# Impacts of nature deprivations during the COVID-19 pandemic: A pre-post comparison

Agathe COLLEONY [1], Susan CLAYTON [2] & Assaf SHWARTZ [1]

[1]Human and Biodiversity Research Lab, Faculty of Architecture and Town Planning, Technion – Israel Institute of Technology, Haifa, 32000 Israel

[2]Department of Psychology, College of Wooster, 1189 Beall Avenue, Wooster, OH 44691, United States

*Corresponding author: Agathe Colléony (agathe.colleony@gmail.com)

Susan Clayton: SCLAYTON@wooster.edu

Assaf Shwartz: shwartza@technion.ac.il

**Highlights**

- We used COVID-19 restriction measures to quantify effects of nature deprivation
- We surveyed the same individuals before the pandemic and during the lockdown
- Experiences of nature dropped for all respondents
- Well-being decreased only for people living in the least green neighborhoods
- Affinity towards nature and environmental attitudes remained unchanged

## Abstract

Nature provides a myriad of intangible and non-material services to people. However, urbanites are increasingly disconnected from the natural world. The consequences of this progressive disconnection from nature remain difficult to measure as this process is slow and long-term monitoring or large-scale manipulation on nature experiences are scarce. Measures to contain the spread of the recent COVID-19 pandemic (i.e., lockdowns) have potentially reduced or even suppressed nature experiences in cities. This situation provided an opportunity for conducting a longitudinal study that can serve as a sort of natural experiment to quantify the effects of nature deprivation on individuals' health, well-being and relationship to nature. We collected data on these variables from the same individuals inhabiting a large metropolis (Tel Aviv, Israel) twice, in 2018 (before) and during the lockdown in 2020. Our results confirmed that frequency, duration and quality of nature interactions dropped during the lockdown, while environmental attitudes and affinity towards nature remained similar. This was particularly true for people living in the least green neighborhoods, where a significant decrease in personal and social well-being was also found. Finally, affinity towards nature influenced well-being through nature experiences in 2018. The mediation effect was not significant in 2020, probably due to the decrease in nature experiences during the lockdown, but the direct relationship between affinity towards nature and well-being remained strong. These results provide insights into the means required to align the public health and conservation agendas to safeguard urbanites' health and well-being during a pandemic and mitigate the biodiversity crisis.

**Keywords**: extinction of experience; nature relatedness; nature experiences; urban green spaces; public health.

# Introduction

Nature experiences contribute to the well-being of individuals, and this is particularly true in cities (Aronson et al., 2017; Soga and Gaston, 2016). These experiences also contribute to developing care for the natural world, which can indirectly help mitigate the biodiversity crisis (Clayton and Myers, 2010). However, increasing urbanization and urban lifestyles have progressively disconnected individuals from nature in a process called the 'extinction of experience' (Soga and Gaston, 2016). This pervasive process threatens humanity's well-being and may aggravate the biodiversity crisis. Quantifying the deleterious effects of the extinction of experience is challenging, as this process is slow and gradual, and long-term monitoring of nature interactions is scarce. The recent COVID-19 pandemic provided an opportunity to assess the consequences of nature deprivation on health, well-being, and relationships with nature (Soga et al., 2021).

Local or national lockdowns were broadly implemented, at least in some countries, to contain the spread of the virus and *flatten the curve*. Because individuals were encouraged or forced to stay home or within a close range to their home, the lockdowns most likely diminished or even suppressed opportunities to experience nature (Day, 2020; Soga et al., 2021). In some cases, illness or forced isolation may have prevented people from visiting nearby nature. Alternatively, adoption of remote working policies may have increased some individuals' available time for nature interactions in their neighborhood (Derks et al., 2020; Soga et al., 2021). At the same time, it is also possible that motivation to interact with nature was affected; there is some evidence that the importance of nature experiences increased during the lockdowns (da Schio et al., 2021; Rousseau and Deschacht, 2020). In addition to causing nearly a million human deaths worldwide, the pandemic has diminished the health and well-being of many more individuals. An increase in nature interactions during the pandemic could help mitigate some of the deleterious effects of the pandemic on mental health. However, a decrease or even deprivation from nature experiences during the pandemic could largely exacerbate the threat to human health.

Opportunities to experience nature and connection to nature are the main drivers of nature experiences (Lin et al., 2014). Urban green spaces provide opportunities for individuals to experience nature close to their home. They can therefore contribute to urban dwellers' health and well-being and promote care for the natural world. High quality nature interactions (e.g., smell flowers, observe wildlife) are associated with increased well-being benefits, connection to nature, ecological knowledge and conservation behaviors (Colléony et al., 2020b; Prévot et al., 2018). Understanding the changes in human interactions with nature during the COVID-19 lockdowns and how these may have influenced individuals' well-being and relationship with nature (connection to nature, environmental attitudes and behaviors) can help identify means to align public health and conservation agendas for a sustainable future. Recent research efforts have focused on exploring such changes in nature interactions during the pandemic. However, existing studies either relied on perceived changes in nature interactions (Grima et al., 2020; Randler et al., 2020) or measured changes through indirect data on green spaces visitation (e.g., mobile phone geolocation) (Day, 2020; Derks et al., 2020). To date, no longitudinal study, to our knowledge, has compared experiences of nature before and during the lockdowns, among the same individuals. Similarly, we are not aware of any study that quantified changes in experiences of nature along with other factors, such as health or well-being. Although recent COVID-19 studies provided us insights on changes in nature interactions during the pandemic (e.g., da Schio et al., 2021; Rousseau and Deschacht, 2020) , our understanding of the extent to which changes in nature interactions during the pandemic affected human health remains very limited.

In this study, we aimed to bridge this knowledge gap, conducting a longitudinal study in which we quantified changes in urban nature experiences, health, well-being, environmental attitudes and behaviors before and during the COVID-19 lockdowns, among the same individuals. Specifically, we replicated a previous survey aiming to understand the complex network of relationships driving the extinction of experience and explored: (1) how individuals perceived biodiversity and green spaces during the COVID-19 lockdown; (2) how nature experiences changed during the lockdown; and (3) what were the changes associated with the deprivation from nature due to the lockdown in individuals' health, well-being and relationship with nature.

**Methods**

Study design

We conducted a survey exploring nature experiences, individual health and well-being, connection to nature and environmental attitudes and behaviors. The same individuals answered the survey twice, during spring 2018 and during the lockdown (spring 2020). The 2018 survey comprised 523 adult residents of Tel Aviv, living along a gradient of urban development (i.e. green, moderately green, and grey neighborhoods; see Colléony et al., 2020a). In 2020, we replicated this survey by reaching out to the individuals who answered the 2018 survey: 325 respondents of the 2018 survey consented to answer the survey in 2020. Of these 325 respondents, 219 were living in the same place as in 2018 (i.e. they did not move). In 2018 and 2020, the survey was administered during the same time of the year and through a market research company (iPanel), which offered respondents monetary compensation for their time loss in return for participation. The questionnaire was distributed in Hebrew and permission for this survey was granted by the Technion Social and Behavioral Sciences Institutional Review Board (approval number: 2018-025). No personal information (e.g., name, contact information) was recorded in the surveys and only the market research company was able to contact participants who had previously registered to their services. Each participant was identified by a unique ID number that was identical for both surveys. All participants were provided with a brief description of the study and gave informed consent for participation.

Questionnaire design

We strictly replicated Colléony et al. (2020a) by recording the following variables:

*Opportunity and orientation*

We measured opportunity to experience nature through measures of nature exposure and greenspace access around participants' address (see Colléony et al., 2020a for details). We measured urban nature exposure using respondents' approximate address and spatial analyses: we calculated an average Normalized Difference Vegetation Index (NDVI) score, the number of trees and the green space area within a 250m buffer around each respondents' home. Greenspace access was recorded by asking respondents to report the average time (0-120 minutes) it takes them to walk to the closest urban green space (urban park, public garden, sidewalk) and reverse-coding values so that a low score of time to walk (e.g., 5min) becomes a high score (in this case, 115) of reported proximity to urban green spaces (see Colléony et al., 2020a for details).

We measured orientation to nature using the short version of the Nature Relatedness Scale (NR6; Nisbet and Zelenski, 2013), designed to capture individuals' affinity towards nature. Inter-item reliability was high (mean score; Cronbach's alphas $\alpha_{2018}$ = 0.88 and $\alpha_{2020}$=0.87).

*Current and childhood experiences of nature*

We measured the quantity of current experiences of nature by asking respondents to estimate the average number of days per month (0 to 30) they visit urban green spaces during the spring. As a measure of duration of visits, we asked respondents to estimate the average duration of each visit to urban green spaces, in a scale ranging from 0 to 420 minutes (7 hours). To assess the quality of nature interactions we followed Colleony et al. (2020b) and recorded the extent to which, on average, participants perform different nature-related behaviors (e.g., watch animals, observe flowers) during their visits to green spaces during Spring. Inter-item reliability was high (mean score; $\alpha_{2018}$ = 0.88 and $\alpha_{2020}$=0.88).

For childhood experiences, we asked respondents to report the average time (0-120 minutes) it used to take them to reach different types of open green spaces during their childhood (6-12 years old). We reverse-coded the values as presented in the *opportunity* section and reported for each participant the value of proximity to the nearest open green space as a single measure of *proximity to open green space during childhood,* a proxy for childhood experiences of nature (see Colléony et al., 2020a for details).

*Health and well-being*

We measured health with two different variables. First, we used the short version of the Depression, Anxiety and Stress Scale (DASS-21). Inter-item reliability ($\alpha_{2018}$ = 0.89 and $\alpha_{2020}$=0.91) was high, so we reversed the scores of each item and summed them to derive a single positive measure of *depression,* with high score for respondents who have low depression, i.e. better mental health. We also measured stress, with the Perceived Stress Scale (Cohen et al., 1983). Inter-item reliability was high ($\alpha_{2018}$ = 0.86 and $\alpha_{2020}$=0.88). Items were reversed and summed to derive a single positive measure of *stress*, with high score for respondents with low level of stress, i.e. better mental health. Finally, we also asked respondents to report their weight and height, for *Body Mass Index* (BMI) calculation (BMI=body mass/(body height)$^2$), and the number of days they performed a physical activity for more than 30 minutes during the past week, as a measure of *physical activity*. We built an index of BMI ranging from 1 poor to 3 good health condition, attributing 1 to obese individuals (BMI > 30), 2 to individuals underweight (BMI < 18.5) and those overweight (25 < BMI < 29.9) and 3 to individuals with normal weight (18.5 < BMI < 24.9) (CDC, 2021). Some studies have shown that self-reported BMI can be lower than the actual values (e.g., Roberts, 1995), yet sufficiently accurate for epidemiological studies (McAdams et al., 2007). While our BMI values might be slightly reduced, we expect these differences to be consistent between the surveys, as we used the same method before and during the lockdown surveys, and therefore this bias should not affect our results.

Following Luck et al. (2011), we used two different scales to measure subjective personal (PWB) and neighborhood well-being (NWB). PWB scale consists of nine items that represent different aspects of overall satisfaction with one's life. NWB scale consists of nine items that represents residents' level of satisfaction with living in their neighborhood. Inter-item reliability was high for both PWB and NWB (mean scores; PWB: $\alpha_{2018}$ = 0.88 and $\alpha_{2020}$=0.90; NWB: $\alpha_{2018}$ = 0.94 and $\alpha_{2020}$=0.94).

*Environmental attitudes and behaviors*

Environmental attitudes were assessed through the 5-item reduced version (Stern et al., 1999) of the New Ecological Paradigm (Dunlap et al., 2000). Inter-item reliability was moderate (mean scores; $\alpha_{2018}$ = 0.60 and $\alpha_{2020}$=0.68). We measured conservation behaviors based on Cooper et al. (2015), with a subscale assessing environmental lifestyle behaviors (3 items; e.g., 'I recycle paper, plastic, metal') and another assessing conservation behaviors (6 items; e.g., 'I made my yard or my land more desirable for wildlife'). Inter-item reliability was high for environmental lifestyle and conservation behaviors (mean scores; environmental lifestyle: $\alpha_{2018}$ = 0.74 and $\alpha_{2020}$=0.80; conservation: $\alpha_{2018}$ = 0.82 and $\alpha_{2020}$=0.85) (see Colléony et al., 2020a for details).

Demographics

Both surveys recorded perceived income. We reported that the average monthly income per household in Israel is 15,000NIS, and asked each participant to rate, from 0 for low to 10 for high, their own household's income (following Shwartz et al., 2012). In both surveys, based on participants' address, we recorded the socioeconomic status of the statistical area they live in (10-point scale, from 1 – low to 10 – high socioeconomic status (Tel Aviv GIS Department, 2018). The 2018 survey recorded age, education, and gender. The 2020 survey recorded the number of respondents who have been tested positive or negative for COVID-19, and the number of respondents who have not been tested but showed symptoms.

*Additional variable: perception of biodiversity and green spaces during the lockdown*

In addition to strictly replicating Colléony et al. (2020a), we developed a novel set of 14 statements assessing individuals' perceptions of biodiversity and green spaces during the lockdown due to the COVID-19 pandemic (Fig. 1). Respondents were asked to report the extent to which they agree to each statement (1 strongly disagree to 5 strongly agree).

Statistical analyses

All spatial analyses were done using ArcGIS 10.5.1, and statistical analyses using R 3.6.0 (R Core Team, 2013). We first checked and confirmed that there were no socioeconomic differences between the three types of neighborhoods (green, moderately green, and grey) (Kruskal-Wallis test; $\chi^2$=4.32, df = 2, p=0.11). We also confirmed that income did not change from 2018 to 2020 (Wilcoxon paired test; V=3947, p=0.15). We conducted a factor analysis on the variables of perception of biodiversity and green spaces during the lockdown: six items loaded together on one factor (understood importance of access to nature – 0.60, discovered green around home – 0.60, enjoyed green around home – 0.81, visiting green around my home contributed to my well-being – 0.85, enjoyed hearing birds – 0.58, noticed more biodiversity – 0.59), two items loaded on a second factor (prevent animals in cities – 0.70, bothered to see animals – 0.78) and two items loaded on a third factor (green around home is poor – 0.55, not happy green around home – 0.51) while the four remaining items did not load on any factor.

We looked at respondents' perception of biodiversity and greenspaces during the lockdown, based on all responses from the 2020 survey (N=325). We then explored changes for each variable between the two points in time for the three different types of neighborhoods (i.e., green, moderately green, and grey) using paired Wilcoxon tests as variables were not distributed normally, based on data from respondents who were in similar conditions (i.e., did not move, similar income) in 2018 and 2020 (N=219). We controlled for multiple testing with the Benjamini-Hochberg correction.

Finally, we tested the overall network of relationships between opportunity, orientation, experiences of nature, health, well-being, environmental attitudes and behaviors using a Structural Equation Model (SEM), with lavaan package (Rosseel, 2012) using the 2020 survey data and compared the resulting model to the same SEM from 2018 (Colléony et al., 2020a). For the 2020 SEM analysis, we used the data from the 219 respondents who were in similar conditions in both surveys. Conventionally considered fit indices in SEM literature have been taken into account to assess the model fit, such as the Root Mean Square Error of Approximation (RMSEA), the Comparative Fit Index (CFI) and the Standardized Root Mean Square Residual (SRMR) (Schreiber et al., 2006). Because data did not follow normal distribution, we used maximum likelihood (ML) for estimating the model parameters, robust standard errors based on a sandwich-type covariance matrix and the Satorra-Bentler scaled test statistic to correct the model test statistics (Rosseel, 2012).

We used each high-level variable as latent constructs (Fig. 4): for instance, health was entered in the model as a variation combining the score of depression, perceived stress, BMI and physical activity. Demographics (i.e. age, income, education, gender) were included as covariates to demonstrate that the predicted relationships were not driven by sociodemographic differences.

# Results

Respondents who took part in both surveys (N=345) were 60% female, on average 42±13 years old, and lived in green (28%), moderately green (36%) or grey (36%) neighborhoods of Tel-Aviv. The vast majority (N=317) were not tested and showed no symptoms of COVID-19. Overall, people showed appreciation for nature's role during the lockdown. Most individuals were glad that nature was repairing itself in the absence of human action, enjoyed hearing more birds and became more aware of the importance of access to nature close to their home during the COVID-19 crisis (Fig. 1). About half of the respondents reported that the green spaces around their home contributed to their well-being (Fig. 1). Only few respondents were not satisfied with the quality of the green spaces close to their home or showed fear of nature or animals in the city during the lockdown (Fig. 1).

Comparing individuals who lived in the same apartment or house in 2018 and 2020 (N=219, 29% in green, 39% in moderately green and 32% in grey neighborhoods), we found significant changes in their experiences of nature. Duration of visits to urban parks and the extent to which individuals interacted with nature significantly decreased during the lockdown for all individuals (Fig. 2b-c; Table S3). While the frequency of visits to urban parks did not change for individuals living in green neighborhoods, it significantly decreased for those in moderately green and grey neighborhoods (Fig. 2a; Table S3).

Average scores of perceived stress, personal and neighborhood well-being, nature relatedness, environmental attitudes and conservation behaviors did not significantly change across time (Fig. 3; Table S3), while the extent to which participants reported depression significantly increased in all neighborhoods (Fig.3; Table S3). All respondents reported more physical activity during the lockdown in 2020 than at the same period in 2018 (Fig. 2d; Table S3). However, looking at the individual items of each scale for each neighborhood type, we found some evidence for deterioration of mental health and well-being, and changes in relationship with nature, for some individuals. Respondents in grey neighborhoods were less satisfied with their standard of living, their health and with what they are achieving in life during the pandemic than in 2018; these respondents also reported being less happy in their neighborhood and less satisfied with the opportunities for rest and relaxation in their neighborhood during the pandemic than in 2018 (Fig. 3; Table S3).

The appeal of remote, wilderness areas diminished during the pandemic compared to 2018 for respondents living in moderately green or grey neighborhoods (Fig. 3; Table S3). Those in green areas reported taking less notice of wildlife during the pandemic than in 2018 (Fig. 3; Table S3). None of the items measuring environmental attitudes changed over time. Finally, we noted significant decreases over time in environmental behaviors for individuals in green neighborhoods and in conservation behaviors for individuals living in grey neighborhoods (Fig. 3; Table S3).

Replication of the previously-observed structural equation model exploring the relationships between drivers and outcomes of nature experiences in 2018 (Colléony et al., 2020a) revealed some differences in the relationships between opportunity, orientation (nature relatedness) and outcomes in 2020 during the lockdown (Fig. 5-6). Opportunity (access), which was positively related to health and well-being in the 2018 survey, was linked to health, well-being, environmental attitudes and behaviors. During the lockdown, nature relatedness was positively associated with experiences of nature, well-being, environmental attitudes and behaviors. However, the relationships between nature experiences and outcomes (well-being and conservation behaviors) which was observed in 2018 disappeared in 2020 (Fig. 5-6). Importantly, nature experiences mediated the relationship between nature relatedness and well-being and the relationship between nature relatedness and conservation behaviors in 2018. Conversely, there were direct relationships between nature relatedness and well-being, between nature relatedness and environmental attitudes and between nature relatedness and behaviors in 2020 (no mediation). Finally, among other demographic variables we accounted for, income was positively correlated with health and well-being outcomes in the two surveys, but the relationship between income, opportunity and orientation was significant only in the before survey in 2018.

## Discussion

While nature could help maintain the health and well-being benefits of urban residents during a crisis such as the COVID-19 pandemic, efforts to contain the spread of the virus paradoxically further diminished individuals' opportunities to experience nature in cities (Day, 2020; Kleinschroth and Kowarik, 2020; Soga et al., 2020; Ugolini et al., 2020). Here, we provide evidence of the reduction in nature experiences due to the COVID-19 lockdowns and the consequences for individuals' health, well-being and relationship with nature. Our longitudinal survey, which can be considered as a natural experiment for nature deprivation, reveals that the network of relationships between opportunity to experience nature, orientation (nature relatedness), experience of nature and various outcomes changed during the lock down. Furthermore, we showed contrasting effects for individuals living in green neighborhoods and those living in grey neighborhoods. Thus, nature deprivation could have stronger effect for those living in grey neighborhoods where opportunity to experience nature are scarce.

The COVID-19 pandemic largely modified our interactions with nature (Soga et al., 2021). While previous studies found both increases and decreases (e.g., Day, 2020; Derks et al., 2020) in nature interactions during the pandemic, our repeated surveys among the same individuals provided empirical evidence of a large decrease in urban nature experiences during the COVID-19 lockdown. Contrasting results across studies may be due to cultural differences in human-nature relationships, as people use and value nature differently across cultures (Colléony et al., 2019). The type of data collected may also explain the differences, as many studies relied on self-reports of perceived changes (e.g., Grima et al., 2020; Randler et al., 2020), which can be influenced by social norms that vary with cultures, unlike more objective measures such as green spaces visitation data or before/after comparisons as we conducted here. Finally, differences in lockdown policies (e.g., perimeter allowed for visits around individuals' home) and green space access and exposure across countries (Kabisch et al., 2016) are other potential factors driving these differences. Future research can seek to compare the consequences of different policies in different countries on human-nature interactions and the associated health and well-being benefits to enable profound understanding of the topic and promote effective policies. In our study, decrease in urban nature experiences was the strongest in grey neighborhoods. This result is not surprising given the important role of opportunity in driving nature interactions (Lin et al., 2014). The issue is concerning because it demonstrates that lockdown policies did not equally affect individuals regarding nature experiences, potentially leading to further inequalities in associated outcomes for health and well-being.

Our study provided empirical evidence that deprivation from nature experiences can alter the network of relationships driving the extinction of experience. The covid-19 pandemic and lockdowns have impacted various aspects of people lives including concerns for financial security, social connectedness, and nature deprivation (Miklitz et al., 2021; Nitschke et al., 2021; Soga et al., 2020). Although we cannot distinguish between various impacts of the lockdown, the fact that we found differences for the same people in the network of relationships associated with experience of nature could shed light on the importance of these nature experiences. Before the lockdown we recorded a significant set of relationships between the frequency, duration and quality of nature experience on one side, and well-being, environmental and conservation behaviors on the other (Colléony et al., 2020a). We also found that experience of nature mediated the relationship between orientation (nature relatedness), well-being, environmental attitudes and conservation behavior. As we expected, during the lockdown the experience of nature was limited due to the regulations and in turn the direct and indirect effects of nature experience on outcomes did not persist. Instead, opportunity to experience nature, measured as proximity to urban parks, remained associated with health and well-being and became associated with environmental and conservation attitudes and behaviors. Thus, our results are consistent with other studies which showed that people who lives closer to urban parks or green spaces have higher health and well-being outcomes (e.g., White et al., 2013). In our case, this was found to be regardless to the frequency, duration, and quality of nature interaction (during the lockdown) and after controlling for socio-demographic variables, including income.

These results were related with our findings that only individuals living in nature impoverished neighborhoods (i.e., grey) demonstrated reduction in some aspect of personal and neighborhood well-being. All respondents in our surveys acknowledged the importance of green spaces for their well-being, but well-being decreased only for individuals who were in the least green neighborhoods. Depression and physical activity changed more uniformly during the lockdown, regardless of the level of greenery in the neighborhood. These contrasting results suggest that reductions in nature experiences may differently affect health and well-being for people who are differently situated, regardless of their socioeconomic status. Changes in depression and physical activity may be due to other factors related to the pandemic rather than a change in nature experiences since we found no difference across neighborhood types. The different changes in well-being across neighborhood types suggest that access to nature was an important factor driving well-being during the pandemic, beyond any impacts of physical activity. Socio-economic differences are unlikely to have driven such changes, as we found no significant difference in socio-economic levels between the three neighborhood types. Our findings therefore emphasize that the large inequalities in access to green space further translate in inequalities regarding health and well-being (Aronson et al., 2017). Recognizing the important role that urban nature can play in health, conserving biodiversity in cities can help to provide more opportunities to experience nature and, by facilitating connection to nature, promote the wellbeing of those who are most deprived from nature experiences.

Interestingly, although people appreciated their nature experiences during the lockdown, we did not find significant changes in environmental attitudes. Individuals' sense that nature was repairing itself during the lockdown may explain these results, suggesting that people are confident in the planet's ability to endure or recover from human activities. This lack of increased concern is concerning, given that the climate and biodiversity crises are still rapid and profound, and wholesale behavior change crucial (Amel et al., 2017; Borrelle et al., 2020; IPBES, 2019). Individuals' connection to nature remained largely unchanged as well, except for the appeal for remote and wild place that decreased significantly among individuals who were the most deprived from nature experiences. As people were deprived from both nature and social interaction during the lockdown, they may have desired more social nature experiences. Accordingly, a recent study found that social interactions largely decreased among birders during the pandemic, as people focused mostly on yard birding (Randler et al., 2020).

The temporary deprivation from nature did not affect connection to nature and its role in driving well-being, environmental attitudes, and conservation behaviors. These results are expected, as connection to nature is a stable and enduring trait of individuals that develops mainly during childhood (Chawla, 2020; Clayton and Myers, 2010). However, connection to nature can also change momentarily (i.e., state, as opposed to trait), for instance after a particular experience in nature (Colléony et al., 2020b). The influence of nature relatedness on well-being remained strong even after the pandemic-induced deprivation from nature experiences, suggesting that individuals' trait (i.e., long-lasting) connection to nature is directly linked to their well-being. Colléony et al. (2020b) proposed that the relationship between state connection to nature and well-being is mediated by nature interactions, but nature experiences did not mediate the relationship between connection to nature and well-being in our 2020 SEM results, which is not surprising as individuals were largely deprived from nature experiences during the COVID-19 lockdowns. These results emphasize the importance to well-being of promoting connection to nature by providing children with opportunities for high quality nature interactions at an early age. Long term monitoring of nature interactions and their outcomes for health and well-being however remains crucial to determine potential deleterious long lasting effects of extended period of loss of nature interactions (Colléony et al., 2020a; Soga and Gaston, 2016). Finally, the decreases that we identified in environmental behaviors in green neighborhoods, and in conservation behaviors in grey neighborhoods are potentially due to constraints on behavior associated with the lockdown.

This study has several limitations. First, the number of individuals who moved within two years, and thus who were not included in the before-during comparison analysis, was surprisingly high. The lower number of participants in the second survey likely diminished the strength of the 2020 structural equation model. Second, as this is a correlational survey, we cannot determine the cause-and-effect relationship between the variables. Although theoretical models consider the effects of nature experiences on health and well-being (Colléony et al., 2020a; Soga and Gaston, 2016), there may be a reverse causality, especially during a pandemic, health and well-being conditions potentially shaping nature experiences. Similarly, we did not examine how changes in economic conditions of participants due to the pandemic have affected their health/well-being and relationship to nature. A number of participants may have lost their job during the pandemic, putting them in very precarious situation that may have affected their health and well-being regardless of nature interactions. However, we did not find support for this as there was no significant change in participants' income across time. Finally, this study relies on self-reports (e.g., for BMI), and self-reports can be biased (e.g., social desirability). However, the strength of this study is that it is longitudinal and compares data from the same individuals. There is no reason to believe self-report bias have changed between surveys. If there were any bias in the answers, these biases should remain consistent and therefore do not undermine the analysis. This is unlike many other COVID-focused studies (e.g., Randler et al., 2020; Soga et al., 2020) that build on data collected only during the pandemic.

## Conclusion

Using the COVID-19 lockdown as a sort of natural experiment in human-nature experiences (Soga et al., 2021), we showed for the first time the direct effects of nature deprivation on health, well-being and relationship with nature. We demonstrated that experience of nature was reduced for all participants during the lockdown, and that some personal and neighborhood well-being aspects were lower for individuals living in areas which were nature poor (i.e., grey neighborhood). We also show that in the absence of nature experiences living closer to green spaces can contribute to health and well-being. Access to nature should be available to all, including urban dwellers; in areas where nature is least present, such as urban areas, efforts should be invested in promoting connection to nature. This can help enhance residents' quality of nature interactions, promote their well-being, and foster care for the natural world, ultimately helping to align public health and conservation agendas (Colléony et al., 2020b; Colléony and Shwartz, 2019; Prévot et al., 2018). However, the network of relationships between the causes and consequences of the extinction of experience is complex and likely to vary according to the characteristics of a specific culture and locale. Further research can benefit from exploring this network of relationships under different geographical and cultural contexts to more fully understand the role of nature experience on well-being and health.

## Supporting Information

Further details on the variables used in the surveys and Structural Equation Model (Tables S1-S2), on the statistics of the pre-post comparison (Table S3), and the survey material can be found in Supporting Information.

## Acknowledgements

This project was funded by the German Israel Foundation (Grant No. I-2490-115.4/2017) and AS is supported by the European Research Council (ERC) under the European Union's Horizon 2020 - Research and Innovation Framework Programme (Starting Grant No. 852633). SC benefitted from a FIAS fellowship at the Paris Institute for Advanced Study (France). It received funding from the European Union's Horizon 2020 research and innovation programme under the Marie Skłodowska-Curie grant agreement No 945408, and from the French State programme "Investissements d'avenir", managed by the Agence Nationale de la Recherche (ANR-11-LABX-0027-01 Labex RFIEA+). We thank Masashi Soga and an anonymous reviewer for their insightful comments on the manuscript.

**Figures**

**Figure 1**: Number and proportion (%) of respondents who agreed (green) or disagreed (red) with a list of statements exploring perception of nature and green spaces during the COVID-19 pandemic in 2020, for a sample of Tel-Aviv inhabitants (N=325).

**Figure 2**: Comparisons of (a) frequency (number of days per month) and (b) duration (minutes per visit) of visits to urban parks, (c) nature interactions during nature visits, and (d) number of days of physical activity during the past week, in 2018 before COVID-19 pandemic and in 2020 during the lockdown associated with the COVID-19 pandemic, for a sample of Tel-Aviv inhabitants living in green (dark green), moderately green (light green) or grey (grey) neighborhoods (N=219).

**Figure 3**: Changes in health (perceived stress, depression), well-being (personal and neighborhood well-being), affinity towards nature (nature relatedness), environmental attitudes and environmental and conservation behaviors during the COVID-19 pandemic in 2020 compared to the same period in 2018, for a sample of Tel-Aviv inhabitants living in green, moderately green or grey neighborhoods (N=219). Significant increases are displayed with red arrows going up, and decreases with red arrows going down.

**Figure 4**: Description of the structural equation model tested in 2018 and 2020.

**Figure 5**: Structural Equation Model of the network of relationships between drivers and outcomes of urban nature experiences in 2018 (see (Colléony et al., 2020a)) (N=523). The model fit was considered satisfactory (Robust indices; χ2 = 353.88, df = 155, CFI = 0.90, RMSEA = 0.05, SRMR = 0.05). Arrows represent significant relationships; direct relationships are displayed in black, mediation effects with orange dashed arrows. Estimates (standard errors) and levels of significance (* p<0.05, ** p<0.01, *** p<0.001) are given.

**Figure 6**: Structural Equation Model of the network of relationships between drivers and outcomes of urban nature experiences in 2020 during the lockdown due to the COVID-19 pandemic (N=219). The model fit was considered satisfactory (Robust indices; χ2 = 258.92, df = 143, CFI = 0.87, RMSEA = 0.06, SRMR = 0.07). Arrows represent significant relationships; direct relationships are displayed in black, mediation effects with orange dashed arrows. Estimates (standard errors) and levels of significance (* p<0.05, ** p<0.01, *** p<0.001) are given.

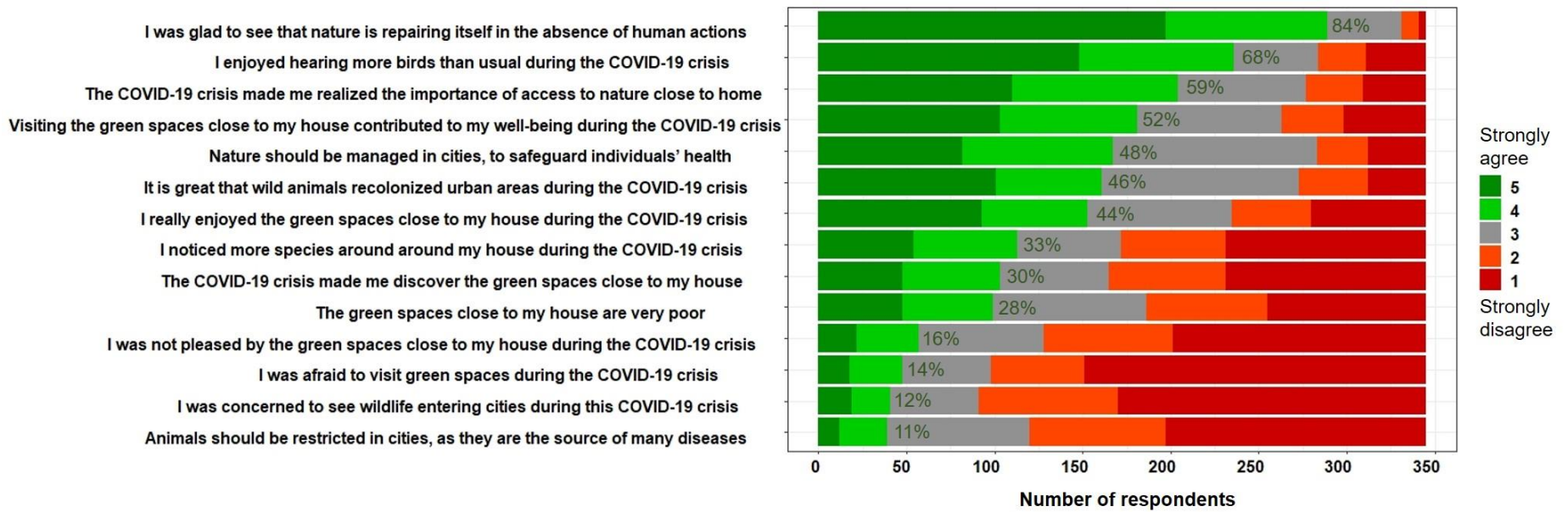

**Figure 1**: Number and proportion (%) of respondents who agreed (green) or disagreed (red) with a list of statements exploring perception of nature and green spaces during the COVID-19 pandemic in 2020, for a sample of Tel-Aviv inhabitants (N=325).

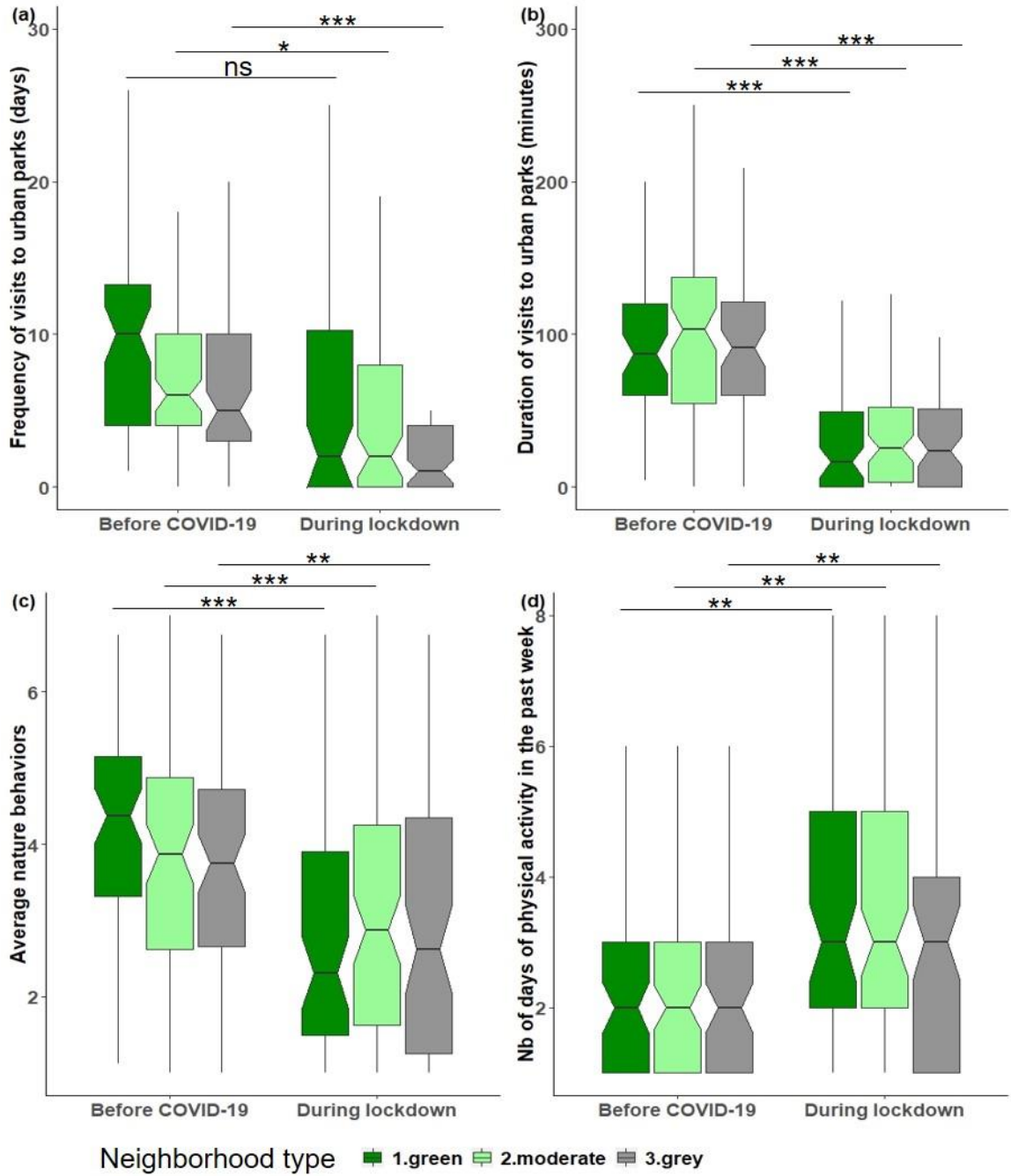

**Figure 2**: Comparisons of (a) frequency (number of days per month) and (b) duration (minutes per visit) of visits to urban parks, (c) nature interactions during nature visits, and (d) number of days of physical activity during the past week, in 2018 before COVID-19 pandemic and in 2020 during the lockdown associated with the COVID-19 pandemic, for a sample of Tel-Aviv inhabitants living in green (dark green), moderately green (light green) or grey (grey) neighborhoods (N=219).

| Measure | Item | Change Green | Change Moderately green | Change Grey | Measure | Item | Change Green | Change Moderately green | Change Grey |
|---|---|---|---|---|---|---|---|---|---|
| Perceived stress | PERCEIVED STRESS (sum) | - | - | - | Personal well-being | PERSONAL WELL-BEING (mean) | - | - | - |
| | I have been upset because of something that happened unexpectedly | ↗ | - | - | | I am satisfied with my life as a whole | - | - | - |
| | I felt that I was unable to control the important things in my life | - | - | - | | I am satisfied with my standard of living | - | - | ↘ |
| | I felt nervous and "stressed" | - | - | - | | I am satisfied with my health | - | - | ↘ |
| | I felt confident about my ability to handle my personal problems | - | - | - | | I am satisfied with what I am achieving in life | - | - | ↘ |
| | I felt that things were going my way | ↘ | - | - | | I am satisfied with my personal relationships | - | - | - |
| | I found that I could not cope with all the things that I had to do | - | - | ↘ | | I am satisfied with how safe I feel | - | - | - |
| | I have been able to control irritations in my life | - | - | - | | I feel part of my community | - | - | - |
| | I felt that I was on top of things | ↘ | - | - | | I am satisfied with my future security | - | - | - |
| | I have been angered because of things that were outside of my control | - | - | - | | I am satisfied with my spirituality or religion | - | - | - |
| | I felt difficulties were piling up so high that I could not overcome them | - | - | - | Neighborhood well-being | NEIGHBORHOOD WELL-BEING (mean) | - | - | - |
| Depression | DEPRESSION (sum) | ↗ | ↗ | ↗ | | I am satisfied with my neighborhood environment as a whole | - | - | - |
| | I couldn't seem to experience any positive feeling at all | ↗ | ↗ | - | | I am satisfied with the opportunity for rest and relaxation here | - | - | ↘ |
| | I found it difficult to work up the initiative to do things | ↗ | ↗ | ↗ | | I feel that this environment reflects who I am | - | - | - |
| | I felt that I had nothing to look forward to | ↗ | - | ↗ | | I am satisfied with the memorable experiences I had here | - | - | - |
| | I felt downhearted and blue | ↗ | ↗ | ↗ | | I am looking forward to spending time here in the future | - | - | - |
| | I was unable to become enthusiastic about anything | ↗ | ↗ | ↗ | | I am able to think about or reflect on personal matters here | - | - | - |
| | I felt I wasn't worth much as a person | ↗ | ↗ | ↗ | | I am satisfied with the advantages of my neighborhood over others | - | - | - |
| | I felt that life was meaningless | ↗ | ↗ | ↗ | | I feel that I belong in my environment | - | - | - |
| Nature relatedness | NATURE RELATEDNESS (mean) | - | - | - | | I feel happy in my neighborhood | - | - | ↘ |
| | My ideal vacation spot would be a remote, wilderness area | - | ↘ | ↘ | Environmental and conservation behaviors | ENVIRONMENTAL BEHAVIORS (mean) | ↘ | - | - |
| | I always think about how my actions affect the environment | - | - | - | | I recycle paper, plastic, metal | - | - | - |
| | My connection to nature and the environment is a part of my spirituality | - | - | - | | I conserve water or energy in my home | ↘ | - | - |
| | I take notice of wildlife wherever I am | ↘ | - | - | | I buy environmentally friendly and/or energy efficient products | - | - | - |
| | My relationship to nature is an important part of who I am | - | - | - | | CONSERVATION BEHAVIORS (mean) | - | - | ↘ |
| | I feel very connected to all living things and the earth | - | - | - | | I make my yard or my land more desirable to wildlife | - | - | ↘ |
| Environmental attitudes | ENVIRONMENTAL ATTITUDES (mean) | - | - | - | | I vote to support a policy or regulation that affects the local environment | - | - | ↘ |
| | I think that the so-called "ecological crisis" facing humankind has been greatly exaggerated | - | - | - | | I donate money to support local environmental protection | - | - | - |
| | I think that the earth is like a spaceship with limited room and resources | - | - | - | | I recruit others to participate in wildlife recreation activities | - | - | - |
| | I think that if things continue on their present course, we will soon experience a major ecological catastrophe | - | - | - | | I volunteer to improve wildlife habitat in my community | - | - | - |
| | I think that the balance of nature is strong enough to cope with impacts of modern industrial nations | - | - | - | | I participate as an active member in an environmental group | - | - | - |
| | I think that humans are severely abusing the environment | - | - | - | | | | | |

**Figure 3**: Changes in health (perceived stress, depression), well-being (personal and neighborhood well-being), affinity towards nature (nature relatedness), environmental attitudes and environmental and conservation behaviors during the COVID-19 pandemic in 2020 compared to the same period in 2018, for a sample of Tel-Aviv inhabitants living in green, moderately green or grey neighborhoods (N=219). Significant increases are displayed with red arrows going up, and decreases with red arrows going down.

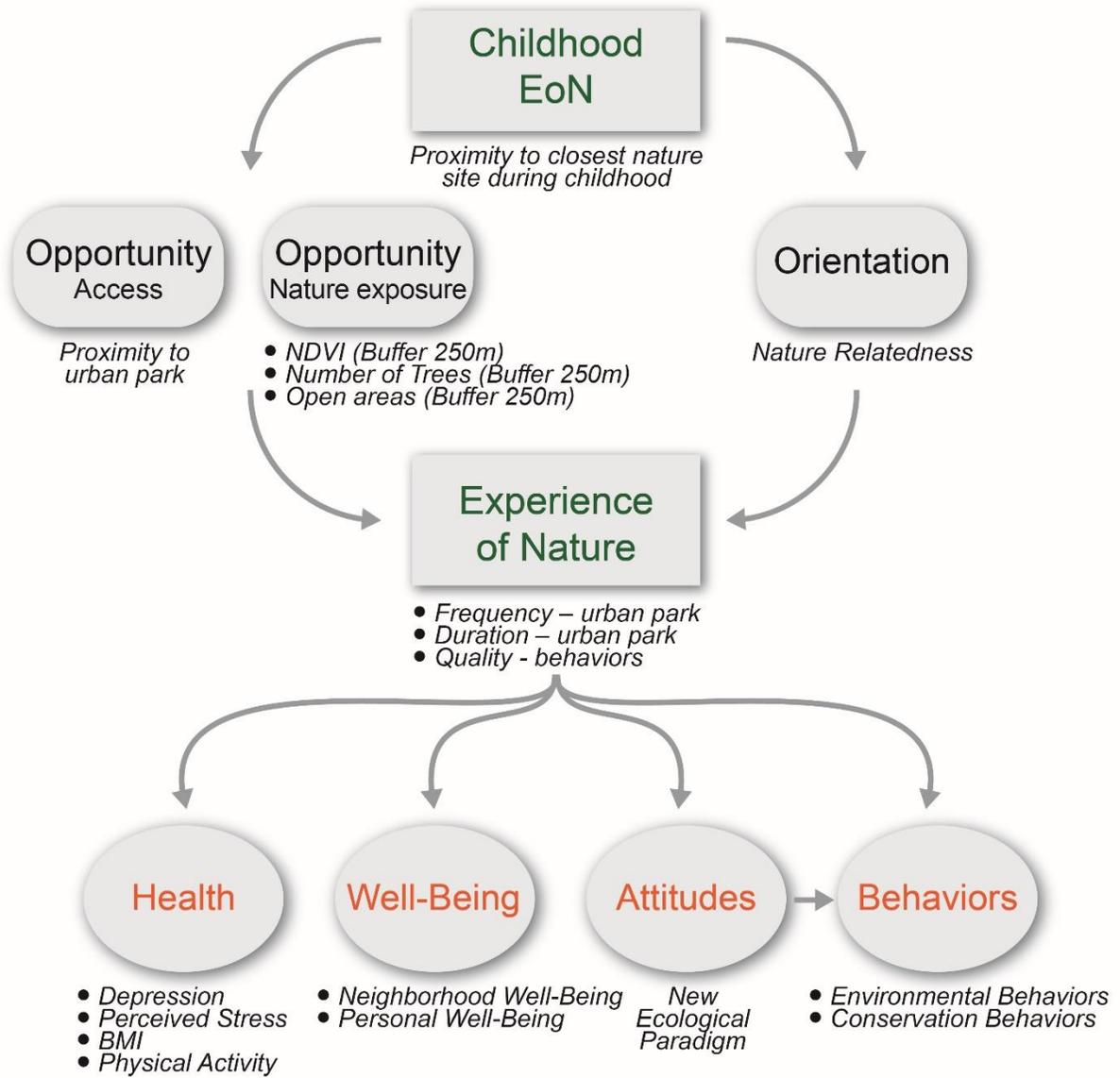

**Figure 4**: Description of the structural equation model tested in 2018 and 2020.

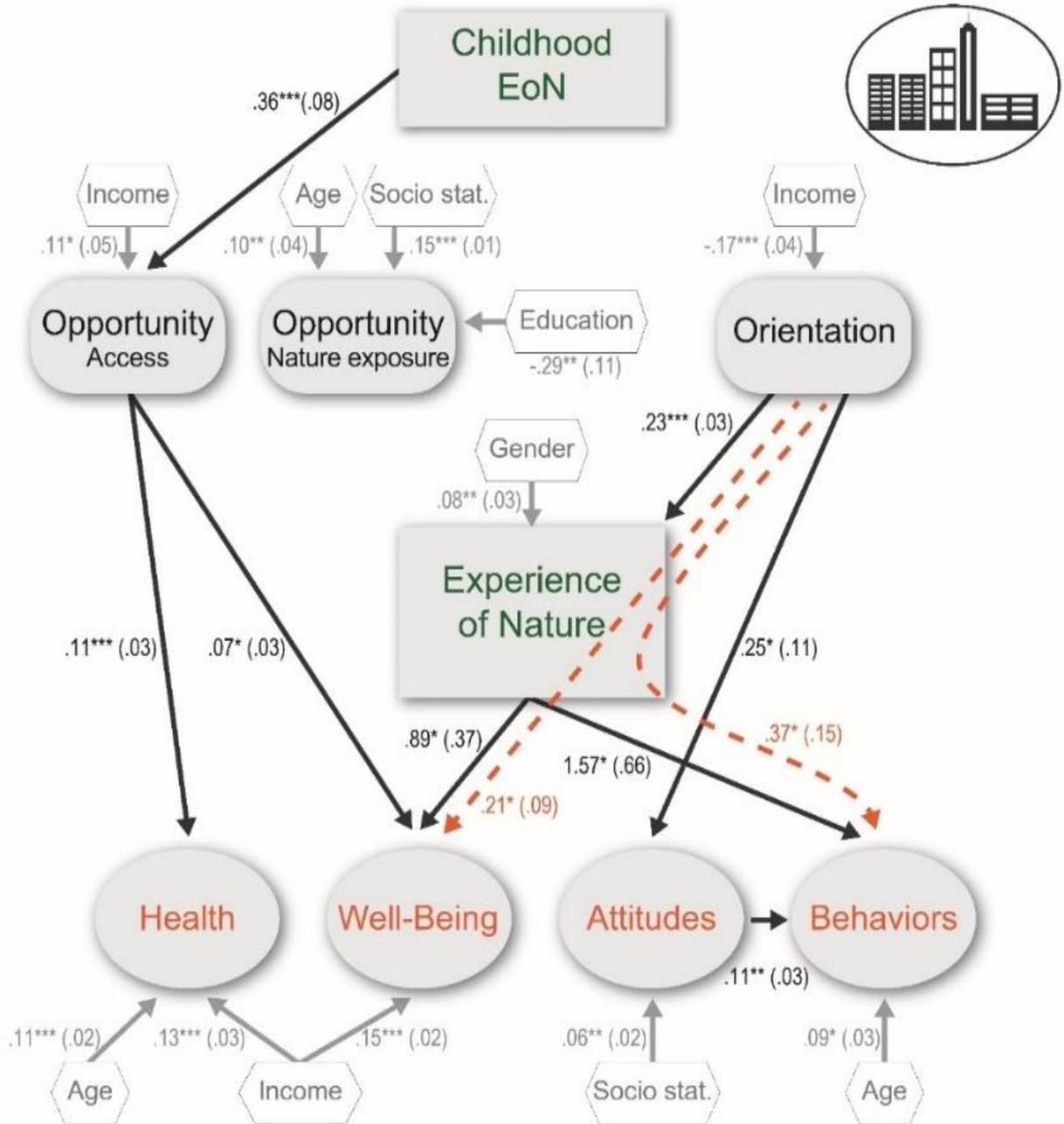

**Figure 5**: Structural Equation Model of the network of relationships between drivers and outcomes of urban nature experiences in 2018 (see (Colléony et al., 2020a)) (N=523). The model fit was considered satisfactory (Robust indices; χ2 = 353.88, df = 155, CFI = 0.90, RMSEA = 0.05, SRMR = 0.05). Arrows represent significant relationships; direct relationships are displayed in black, mediation effects with orange dashed arrows. Estimates (standard errors) and levels of significance (* p<0.05, ** p<0.01, *** p<0.001) are given.

.

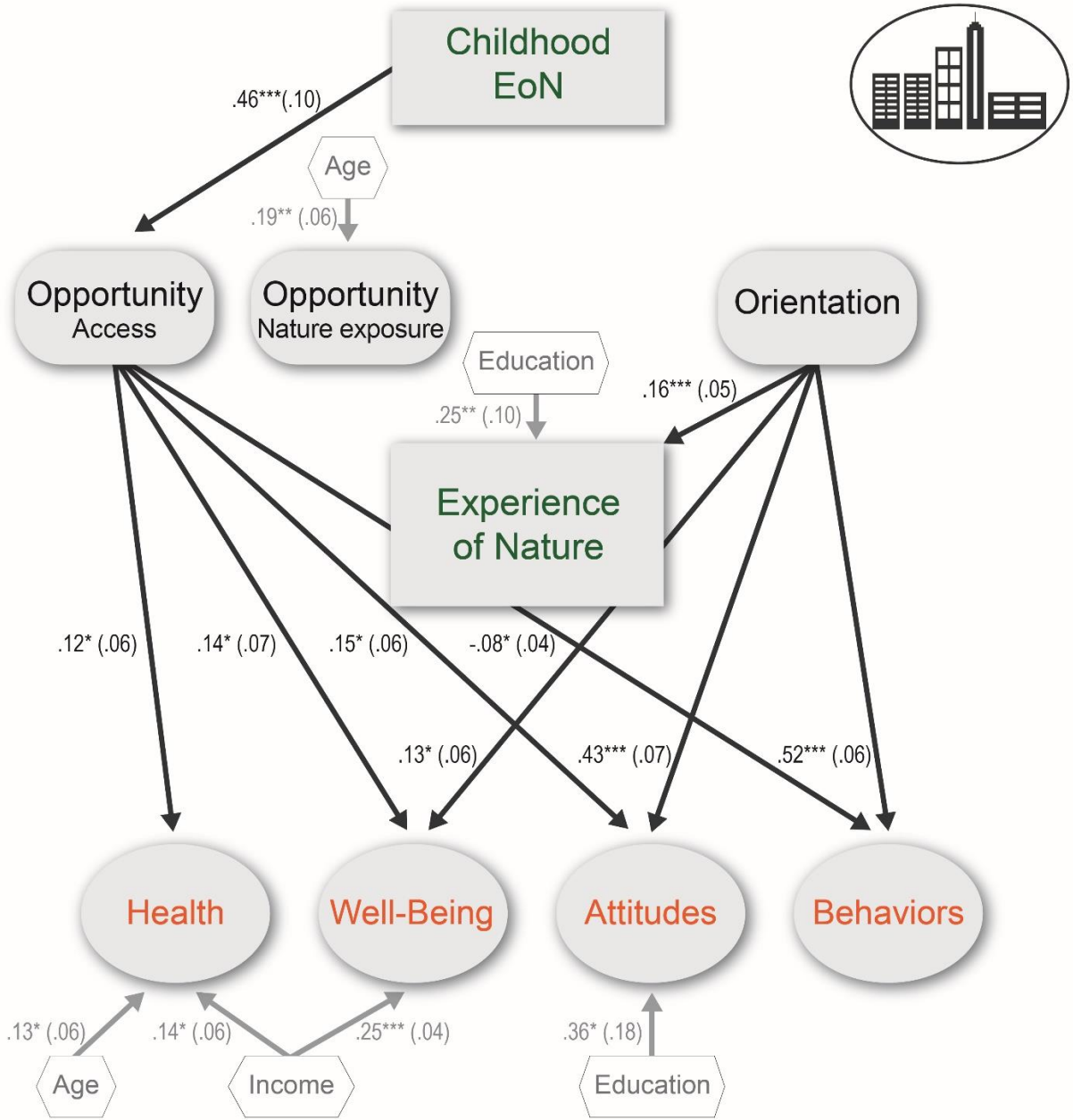

**Figure 6**: Structural Equation Model of the network of relationships between drivers and outcomes of urban nature experiences in 2020 during the lockdown due to the COVID-19 pandemic (N=219). The model fit was considered satisfactory (Robust indices; χ2 = 258.92, df = 143, CFI = 0.87, RMSEA = 0.06, SRMR = 0.07). Arrows represent significant relationships; direct relationships are displayed in black, mediation effects with orange dashed arrows. Estimates (standard errors) and levels of significance (* p<0.05, ** p<0.01, *** p<0.001) are given.

# Supplementary Material

Table S1: Variables that are included in latent constructs for the structural equation model in 2018, their range of values and mean ± standard deviation.

| LATENT CONSTRUCT | VARIABLE | RANGE | MEAN±SD |
|---|---|---|---|
| Opportunity – nature exposure | NDVI Buffer 250m | [-0.02 – 0.34] | 0.19 ± 0.07 |
| | Number of trees Buffer 250m | [0 – 757] | 153.26 ± 103.70 |
| | Green spaces areas ($m^2$) Buffer 250m | [40.11 – 157 880] | 37 154.67 ± 28 131.66 |
| Opportunity – access | Proximity to urban green spaces | [0 – 60] | 51.12 ± 9.18 |
| Childhood experiences of nature | Proximity to open green spaces | [0 – 120] | 110.80 ± 10.49 |
| Orientation | Nature relatedness (Nisbet and Zelenski, 2013) | [1 – 5] | 3.13 ± 1.00 |
| Experience of nature | Frequency of visits to urban green spaces | [0 – 30] | 8.28 ± 6.60 |
| | Duration of visits to urban green spaces | [0 – 420] | 109.46 ± 72.09 |
| | Nature interactions (quality) | [1 – 7] | 3.95 ± 1.38 |
| Health | Depression (Following Shanahan et al., 2016) | [0 – 21] | 17.06 ± 4.60 |
| | Stress (Cohen et al., 1983) | [10 – 50] | 33.20 ± 7.32 |
| | Body Mass Index (index) | [1 – 3] | 2.45 ± 0.69 |
| | Number of days of physical activity in the past week | [1 – 7] | 2.33 ± 1.62 |
| Well-being | Neighborhood well-being (Luck et al., 2011) | [1 – 5] | 3.48 ± 0.93 |
| | Personal well-being (Luck et al., 2011) | [1 – 5] | 3.61 ± 0.74 |
| Environmental attitudes | New Ecological Paradigm (Dunlap et al., 2000) | [1 – 5] | 3.71 ± 0.69 |
| Conservation behaviors | Environmental behaviors (Cooper et al., 2015) | [1 – 5] | 3.67 ± 0.93 |
| | Conservation behaviors (Cooper et al., 2015) | [1 – 5] | 2.01 ± 0.79 |

Table S2: Variables that are included in latent constructs for the structural equation model in 2020, their range of values and mean ± standard deviation.

| LATENT CONSTRUCT | VARIABLE | RANGE | MEAN±SD |
|---|---|---|---|
| Opportunity – nature exposure | NDVI Buffer 250m | [-0.02 – 0.34] | 0.19 ± 0.07 |
| | Number of trees Buffer 250m | [0 – 757] | 153.26 ± 103.70 |
| | Green spaces areas ($m^2$) Buffer 250m | [40.11 – 157 880] | 37 154.67 ± 28 131.66 |
| Opportunity – access | Proximity to urban green spaces | [0 – 60] | 51.12 ± 9.18 |
| Childhood experiences of nature | Proximity to open green spaces | [0 – 120] | 110.80 ± 10.49 |
| Orientation | Nature relatedness (Nisbet and Zelenski, 2013) | [1 – 5] | 3.02 ± 0.95 |
| Experience of nature | Frequency of visits to urban green spaces | [0 – 30] | 5.58 ± 8.73 |
| | Duration of visits to urban green spaces | [0 – 420] | 44.49 ± 67.35 |
| | Nature interactions (quality) | [1 – 7] | 2.90 ± 1.56 |
| Health | Depression (Following Shanahan et al., 2016) | [0 – 21] | 14.61 ± 5.23 |
| | Stress (Cohen et al., 1983) | [10 – 50] | 33.37 ± 7.77 |
| | Body Mass Index (index) | [1 – 3] | 2.22 ± 0.45 |
| | Number of days of physical activity in the past week | [1 – 7] | 3.42 ± 2.16 |
| Well-being | Neighborhood well-being (Luck et al., 2011) | [1 – 5] | 3.51 ± 0.91 |
| | Personal well-being (Luck et al., 2011) | [1 – 5] | 3.53 ± 0.78 |
| Environmental attitudes | New Ecological Paradigm (Dunlap et al., 2000) | [1 – 5] | 3.73 ± 0.71 |
| Conservation behaviors | Environmental behaviors (Cooper et al., 2015) | [1 – 5] | 3.59 ± 0.99 |
| | Conservation behaviors (Cooper et al., 2015) | [1 – 5] | 2.00 ± 0.88 |

**Table S3:** Summary statistics of the paired Wilcoxon tests comparing scores in 2020 to those in 2018 for each variable. Mean and Standard deviation are provided for each variable.

| | | Green neighborhood (N=64) | | | | Moderately green neighborhood (N=85) | | | | Grey neighborhood (N=70) | | | |
|---|---|---|---|---|---|---|---|---|---|---|---|---|---|
| | | V | p-value | Mean±SD 2018 | Mean±SD 2020 | V | p-value | Mean±SD 2018 | Mean±SD 2020 | V | p-value | Mean±SD 2018 | Mean±SD 2020 |
| Frequency visits | - | 1327.5 | 0.176 | 9.53±7.03 | 7.50±10.92 | 2194.5 | 0.0279 | 7.75±6.05 | 5.49±7.75 | 1906.5 | 1.01E-05 | 7.47±6.40 | 3.94±7.25 |
| Duration visit | - | 1800 | 0.000 | 116.90±90.22 | 47.07±82.35 | 3012.5 | 0.0000 | 110.30±76.33 | 45.45±64.43 | 2234 | 6.13E-09 | 105.42±75.03 | 40.95±55.37 |
| Nature behaviors | - | 1684 | 0.000 | 4.18±1.35 | 2.79±1.58 | 2569 | 0.0003 | 3.85±1.42 | 3.02±1.49 | 1798.5 | 0.005178 | 3.71±1.42 | 2.86±1.63 |
| Perceived stress | PERCEIVED STRESS (sum) | 802 | 0.648 | 26.54±7.74 | 27.68±7.24 | 1497 | 0.8975 | 25.83±7.37 | 26.45±7.95 | 1416.5 | 0.1218 | 27.35±6.93 | 25.85±8.02 |
| | I have been upset because of something that happened unexpectedly | 262.5 | 0.079 | 2.50±1.27 | 2.89±1.11 | 683.5 | 0.4873 | 2.43±1.24 | 2.61±1.12 | 589.5 | 0.8457 | 2.60±1.10 | 2.52±1.18 |
| | I felt that I was unable to control the important things in my life | 457 | 0.969 | 2.64±1.28 | 2.67±1.18 | 615 | 0.4243 | 2.47±1.23 | 2.70±1.21 | 452.5 | 0.86 | 2.60±1.23 | 2.60±1.22 |
| | I felt nervous and "stressed" | 320.5 | 0.963 | 3.10±1.11 | 3.12±1.16 | 861.5 | 0.9556 | 3.01±1.17 | 3.03±1.16 | 693 | 0.2871 | 3.00±1.09 | 2.80±1.11 |
| | I felt confident about my ability to handle my personal problems | 327 | 0.822 | 3.46±1.16 | 3.56±1.03 | 658.5 | 0.9847 | 3.65±1.05 | 3.67±1.02 | 485 | 0.6576 | 3.42±1.07 | 3.54±1.12 |
| | I felt that things were going my way | 738 | 0.041 | 3.25±0.95 | 2.84±0.89 | 1082 | 0.2680 | 3.28±0.98 | 3.05±0.98 | 376 | 0.6164 | 3.08±0.97 | 3.20±1.18 |
| | I found that I could not cope with all the things that I had to do | 677 | 0.822 | 2.60±1.19 | 2.53±1.16 | 798.5 | 0.4243 | 2.68±1.18 | 2.44±1.25 | 705.5 | 0.03722 | 2.61±1.10 | 2.24±1.08 |
| | I have been able to control irritations in my life | 338 | 0.670 | 3.40±1.03 | 3.54±0.97 | 565 | 0.9404 | 3.42±1.00 | 3.42±0.94 | 472 | 0.2759 | 3.10±0.98 | 3.32±0.95 |
| | I felt that I was on top of things | 703.5 | 0.045 | 3.51±1.02 | 3.10±1.02 | 705.5 | 0.6142 | 3.56±0.99 | 3.44±1.09 | 395.5 | 0.8794 | 3.25±1.01 | 3.28±1.14 |
| | I have been angered because of things that were outside of my control | 427.5 | 0.763 | 2.90±1.17 | 3.06±1.11 | 1083.5 | 0.7538 | 2.90±1.15 | 2.78±1.22 | 657.5 | 0.46 | 2.88±1.09 | 2.72±1.11 |
| | I felt difficulties were piling up so high that I could not overcome them | 482.5 | 0.963 | 2.42±1.24 | 2.46±1.12 | 627.5 | 0.3608 | 2.25±1.08 | 2.47±1.19 | 626.5 | 0.34 | 2.52±1.23 | 2.31±1.22 |
| Depression | DEPRESSION (sum) | 135.5 | 0.000 | 2.96±3.75 | 7.14±5.52 | 444 | 0.0000 | 3.98±4.77 | 6.87±6.00 | 351 | 0.001224 | 4.28±4.55 | 6.75±5.89 |

| Category | Item | | | | | | | | | | | |
|---|---|---|---|---|---|---|---|---|---|---|---|---|
| | I couldn't seem to experience any positive feeling at all | 150.5 | 0.002 | 1.46±0.68 | 2.01±1.07 | 433 | 0.0183 | 1.60±0.80 | 2.00±1.04 | 298 | 0.2303 | 1.65±0.83 | 1.85±0.98 |
| | I found it difficult to work up the initiative to do things | 72 | 0.000 | 1.70±0.70 | 2.60±1.19 | 369 | 0.0013 | 1.82±0.84 | 2.41±1.26 | 306 | 0.005583 | 1.97±0.85 | 2.47±1.09 |
| | I felt that I had nothing to look forward to | 24 | 0.000 | 1.31±0.63 | 1.92±1.08 | 205 | 0.2062 | 1.47±0.85 | 1.75±1.10 | 99.5 | 0.03722 | 1.50±0.86 | 1.78±1.06 |
| | I felt downhearted and blue | 57 | 0.000 | 1.54±0.71 | 2.45±1.15 | 243.5 | 0.0000 | 1.68±0.87 | 2.49±1.21 | 149 | 0.005583 | 1.71±0.87 | 2.24±1.30 |
| | I was unable to become enthusiastic about anything | 52 | 0.000 | 1.37±0.70 | 2.21±1.07 | 326 | 0.0046 | 1.52±0.74 | 2.02±1.19 | 162 | 0.005178 | 1.65±0.81 | 2.20±1.19 |
| | I felt I wasn't worth much as a person | 15 | 0.000 | 1.29±0.60 | 1.95±1.10 | 158 | 0.0016 | 1.44±0.74 | 1.94±1.12 | 157 | 0.005178 | 1.42±0.67 | 1.94±1.08 |
| | I felt that life was meaningless | 52.5 | 0.000 | 1.26±0.59 | 1.87±1.13 | 192 | 0.0134 | 1.43±0.76 | 1.83±1.17 | 59.5 | 0.005583 | 1.35±0.68 | 1.75±1.08 |
| Physical activity | - | 241 | 0.006 | 2.46±1.77 | 3.56±2.23 | 451.5 | 0.0013 | 2.57±1.74 | 3.52±2.16 | 282 | 0.005583 | 2.35±1.61 | 3.17±2.09 |
| Personal well-being | PERSONAL WELL-BEING (mean) | 1023.5 | 0.830 | 3.59±0.71 | 3.54±0.71 | 1734.5 | 0.9847 | 3.64±0.77 | 3.62±0.78 | 1441 | 0.22 | 3.54±0.73 | 3.40±0.82 |
| | I am satisfied with my life as a whole | 217.5 | 1.000 | 3.71±0.82 | 3.71±0.84 | 312.5 | 0.8070 | 3.87±0.78 | 3.81±0.86 | 309 | 0.2079 | 3.85±0.83 | 3.67±1.08 |
| | I am satisfied with my standard of living | 263 | 0.363 | 3.59±0.93 | 3.45±1.06 | 610 | 0.3608 | 3.83±0.89 | 3.70±1.03 | 413 | 0.09571 | 3.65±0.91 | 3.44±1.04 |
| | I am satisfied with my health | 292 | 0.268 | 3.87±0.84 | 3.70±0.92 | 610.5 | 0.8593 | 3.69±1.01 | 3.63±1.02 | 440 | 0.03608 | 3.77±0.91 | 3.51±1.01 |
| | I am satisfied with what I am achieving in life | 200 | 0.822 | 3.56±1.00 | 3.50±0.95 | 632 | 0.7538 | 3.72±0.96 | 3.64±1.04 | 496 | 0.03722 | 3.64±0.99 | 3.40±1.05 |
| | I am satisfied with my personal relationships | 270 | 0.861 | 3.60±1.12 | 3.86±1.18 | 343 | 0.8794 | 3.62±1.16 | 3.58±1.20 | 397.5 | 0.7817 | 3.54±1.05 | 3.48±1.08 |
| | I am satisfied with how safe I feel | 293 | 0.952 | 3.59±0.95 | 3.57±0.95 | 529 | 0.3350 | 3.77±1.02 | 3.60±0.97 | 541 | 0.1737 | 3.54±0.98 | 3.31±1.14 |
| | I feel part of my community | 402.5 | 0.350 | 3.39±1.07 | 3.21±1.03 | 576.5 | 0.8070 | 3.40±1.06 | 3.47±1.05 | 396 | 0.78 | 3.17±1.02 | 3.11±1.12 |
| | I am satisfied with my future security | 383.5 | 0.852 | 3.25±1.09 | 3.20±1.07 | 442 | 0.8975 | 3.38±1.12 | 3.41±1.04 | 261 | 0.64 | 3.08±1.07 | 3.15±1.13 |
| | I am satisfied with my spirituality or religion | 255 | 0.648 | 3.75±0.95 | 3.85±1.00 | 407.5 | 0.3350 | 3.47±1.14 | 3.76±1.04 | 452.5 | 0.67 | 3.60±1.09 | 3.52±1.15 |
| Neighborhood well-being | NEIGHBORHOOD WELL-BEING (mean) | 871 | 0.963 | 3.61±1.00 | 3.60±0.88 | 1501.5 | 0.6338 | 3.47±0.80 | 3.49±0.95 | 1505 | 0.1737 | 3.62±0.78 | 3.46±0.90 |
| | I am satisfied with my neighborhood environment as a whole | 249.5 | 0.350 | 4.00±085 | 3.82±0.98 | 481.5 | 0.3970 | 3.90±0.88 | 3.75±1.09 | 544 | 0.3518 | 3.98±0.85 | 3.82±1.02 |
| | I am satisfied with the opportunity for rest and relaxation here | 277.5 | 0.952 | 3.85±1.02 | 3.78±1.03 | 376.5 | 0.8975 | 3.64±1.13 | 3.60±1.18 | 608 | 0.09571 | 3.68±0.97 | 3.41±1.09 |

|  | Item | | | | | | | | | | | |
|---|---|---|---|---|---|---|---|---|---|---|---|---|
|  | I feel that this environment reflects who I am | 301 | 0.822 | 3.40±1.29 | 3.31±1.18 | 809 | 0.8189 | 3.32±1.13 | 3.23±1.25 | 563.5 | 0.1737 | 3.58±1.09 | 3.34±1.22 |
|  | I am satisfied with the memorable experiences I had here | 433 | 0.952 | 3.23±1.25 | 3.28±1.16 | 531.5 | 0.2393 | 2.85±1.09 | 3.10±1.15 | 450 | 0.5417 | 3.15±1.08 | 3.04±1.16 |
|  | I am looking forward to spending time here in the future | 347.5 | 0.822 | 3.31±1.35 | 3.42±1.19 | 511.5 | 0.4323 | 3.28±1.04 | 3.43±1.19 | 252 | 0.8721 | 3.47±1.08 | 3.48±1.13 |
|  | I am able to think about or reflect on personal matters here | 460 | 0.925 | 3.45±1.18 | 3.43±1.05 | 389.5 | 0.3608 | 3.30±1.03 | 3.50±1.08 | 426.5 | 0.2436 | 3.45±0.94 | 3.25±1.01 |
|  | I am satisfied with the advantages of my neighborhood over others | 220.5 | 0.830 | 4.01±0.93 | 4.07±0.94 | 652.5 | 0.9784 | 3.91±0.95 | 3.88±1.10 | 384.5 | 0.3759 | 3.91±0.86 | 3.75±1.06 |
|  | I feel that I belong in my environment | 404.5 | 0.946 | 3.54±1.32 | 3.60±1.03 | 732.5 | 0.9556 | 3.48±0.97 | 3.43±1.09 | 472.5 | 0.5243 | 3.58±1.08 | 3.50±1.04 |
|  | I feel happy in my neighborhood | 243 | 0.963 | 3.70±1.10 | 3.71±1.03 | 756 | 0.9726 | 3.51±1.03 | 3.47±1.09 | 520 | 0.03722 | 3.77±0.98 | 3.51±1.01 |
| Nature relatedness | NATURE RELATEDNESS (mean) | 1096.5 | 0.763 | 3.15±1.04 | 3.08±0.89 | 1642 | 0.9258 | 3.05±0.98 | 2.98±0.97 | 1236.5 | 0.4467 | 3.15±1.10 | 3.01±0.98 |
|  | My ideal vacation spot would be a remote, wilderness area | 423 | 0.622 | 3.53±1.18 | 3.40±1.17 | 689 | 0.0043 | 3.50±1.18 | 3.11±1.24 | 569 | 0.03722 | 3.62±1.15 | 3.30±1.15 |
|  | I always think about how my actions affect the environment | 450.5 | 0.952 | 3.34±1.07 | 3.39±1.07 | 659.5 | 0.5715 | 3.40±1.14 | 3.29±1.10 | 573 | 0.22 | 3.30±1.14 | 3.07±1.18 |
|  | My connection to nature and the environment is a part of my spirituality | 503 | 0.472 | 3.10±1.45 | 2.93±1.23 | 518.5 | 0.9404 | 2.91±1.25 | 2.94±1.31 | 463 | 0.91 | 2.84±1.46 | 2.82±1.35 |
|  | I take notice of wildlife wherever I am | 519 | 0.079 | 2.89±1.38 | 2.57±1.15 | 575.5 | 0.4243 | 2.63±1.13 | 2.78±1.23 | 367.5 | 0.5243 | 2.88±1.38 | 2.72±1.27 |
|  | My relationship to nature is an important part of who I am | 342 | 0.822 | 3.06±1.46 | 3.15±1.19 | 554 | 0.8975 | 2.97±1.27 | 2.98±1.32 | 356.5 | 0.6016 | 3.01±1.25 | 3.05±1.31 |
|  | I feel very connected to all living things and the earth | 240.5 | 0.963 | 3.00±1.24 | 3.03±1.14 | 847.5 | 0.7817 | 2.88±1.26 | 2.77±1.15 | 447.5 | 0.5444 | 3.22±1.36 | 3.11±1.30 |
| Environmental attitudes | ENVIRONMENTAL ATTITUDES (mean) | 580 | 0.952 | 3.75±0.70 | 3.80±0.70 | 1032 | 0.9847 | 3.64±0.76 | 3.65±0.73 | 835 | 0.78 | 3.74±0.71 | 3.76±0.68 |
|  | I think that the so-called "ecological crisis" facing humankind has been greatly exaggerated | 231.5 | 1.000 | 2.37±1.11 | 2.39±1.22 | 632 | 0.7538 | 2.54±1.26 | 2.40±1.26 | 417.5 | 0.7817 | 2.38±1.21 | 2.32±1.09 |

| | | | | | | | | | | | | |
|---|---|---|---|---|---|---|---|---|---|---|---|---|
| | I think that the earth is like a spaceship with limited room and resources | 237 | 0.852 | 3.78±1.10 | 3.85±0.94 | 587.5 | 1.0000 | 3.70±1.12 | 3.74±1.05 | 332 | 0.3259 | 3.91±1.00 | 3.75±0.98 |
| | I think that if things continue on their present course, we will soon experience a major ecological catastrophe | 242 | 0.963 | 3.85±1.08 | 3.90±0.93 | 382 | 0.9556 | 3.58±1.14 | 3.55±1.15 | 322 | 0.91 | 3.70±1.20 | 3.68±1.04 |
| | I think that the balance of nature is strong enough to cope with impacts of modern industrial nations | 307 | 0.822 | 2.37±1.06 | 2.46±1.15 | 337.5 | 0.3350 | 2.44±1.13 | 2.67±1.10 | 263 | 0.6578 | 2.34±0.94 | 2.41±1.02 |
| | I think that humans are severely abusing the environment | 209.5 | 0.472 | 3.89±1.14 | 4.09±1.10 | 457 | 0.6142 | 3.91±1.13 | 4.04±0.93 | 118 | 0.1737 | 3.85±1.15 | 4.10±0.90 |
| Environmental and conservation behaviors | ENVIRONMENTAL BEHAVIORS (mean) | 1187 | 0.045 | 3.82±0.85 | 3.61±1.02 | 1741.5 | 0.3608 | 3.67±0.96 | 3.58±0.99 | 1134.5 | 0.3164 | 3.66±0.93 | 3.60±0.99 |
| | I recycle paper, plastic, metal | 110 | 0.963 | 3.89±1.20 | 3.87±1.26 | 415.5 | 0.7817 | 3.90±1.26 | 3.87±1.16 | 256.5 | 0.8968 | 3.74±1.28 | 3.71±1.29 |
| | I conserve water or energy in my home | 462.5 | 0.017 | 4.03±1.03 | 3.62±1.24 | 517.5 | 0.3608 | 3.82±1.11 | 3.64±1.18 | 415.5 | 0.6437 | 3.90±1.06 | 3.82±1.08 |
| | I buy environmentally friendly and/or energy efficient products | 500 | 0.319 | 3.56±1.03 | 3.34±1.11 | 738.5 | 0.8794 | 3.30±1.02 | 3.24±1.07 | 352 | 0.6574 | 3.34±1.15 | 3.25±1.23 |
| | CONSERVATION BEHAVIORS (mean) | 858.5 | 0.822 | 1.96±0.76 | 2.05±0.85 | 1530.5 | 0.8975 | 1.99±0.74 | 2.07±0.95 | 1565.5 | 0.005583 | 2.14±0.96 | 1.87±0.81 |
| | I make my yard or my land more desirable to wildlife | 185.5 | 0.952 | 1.89±1.15 | 1.81±1.23 | 728.5 | 0.3608 | 2.20±1.31 | 2.00±1.30 | 526 | 0.005583 | 2.22±1.22 | 1.74±1.16 |
| | I vote to support a policy or regulation that affects the local environment | 536 | 0.402 | 3.59±1.20 | 3.40±1.16 | 660 | 0.9472 | 3.31±1.13 | 3.30±1.20 | 606 | 0.07053 | 3.41±1.19 | 3.00±1.38 |
| | I donate money to support local environmental protection | 157 | 0.624 | 1.73±1.02 | 1.84±1.15 | 438.5 | 0.4873 | 1.82±1.00 | 1.98±1.24 | 375.5 | 0.1737 | 1.95±1.30 | 1.72±1.04 |
| | I recruit others to participate in wildlife recreation activities | 101.5 | 0.763 | 1.53±0.90 | 1.64±1.01 | 202.5 | 0.9181 | 1.56±0.95 | 1.56±1.02 | 349.5 | 0.2079 | 1.81±1.14 | 1.60±0.95 |
| | I volunteer to improve wildlife habitat in my community | 139 | 0.690 | 1.43±0.85 | 1.59±0.93 | 166 | 0.3350 | 1.49±0.82 | 1.72±1.20 | 208 | 0.34 | 1.64±1.07 | 1.51±0.98 |
| | I participate as an active member in an environmental group | 112 | 0.112 | 1.60±1.07 | 2.01±1.36 | 261 | 0.3350 | 1.58±0.91 | 1.83±1.22 | 281 | 0.4519 | 1.78±1.20 | 1.65±1.10 |

Survey material

**Informed Consent Form**

Dear Participant,

You are invited to take part in a survey that investigate people's health and use of green space.

Your responses will be anonymous. We will only ask you to provide an **approximation** of your address. This information will be stored in a secured file only accessible by the researchers, and will be deleted at the end of the project. There is no risk at taking part of this survey, there is no wrong answer, and information will only be used for research purposes.

The questionnaire will take you about 10 minutes to complete.

Taking part in this study is completely voluntary. If you decide to take part, you are free to withdraw at any time.

Should you have any further questions about this project or if you have a problem of any kind, you may contact the principle supervisor Dr. Agathe Colleony ([agatheco@technion.ac.il](agatheco@technion.ac.il))

By choosing to participate, you are confirming that you are 18 years of age or older, that you have read and understand the information provided above and that you willingly choose to participate in this survey. If you choose to participate, please tick I ACCEPT.

□ I ACCEPT

Thank you!

# Survey of Tel Aviv inhabitants

Have you been tested for corona virus?

- I have been tested positive
- I have been tested negative
- I have not been tested but I experienced related symptoms
- I have not been tested and did not experience any related symptom

In light of the COVID-10 situation, to what extent do you agree with each statement (1 strongly disagree to 5 strongly agree):

|  | 1 Strongly disagree | 2 | 3 | 4 | 5 Strongly agree |
|---|---|---|---|---|---|
| *Animals should be restricted in cities, as they are the source of many diseases* | | | | | |
| *I was concerned to see wildlife entering cities during this COVID-19 crisis* | | | | | |
| *Nature should be limited in cities, to safeguard individuals' health* | | | | | |
| *It is great that wild animals recolonized urban areas during the COVID-19 crisis* | | | | | |
| *I was glad to see that nature is repairing itself in the absence of human actions* | | | | | |
| *The COVID-19 crisis made me realized the importance of access to nature close to home* | | | | | |
| *The COVID-19 crisis made me discover the green spaces close to my house* | | | | | |
| *I really enjoyed the green spaces close to my house during the COVID-19 crisis* | | | | | |
| *The green spaces close to my house are very poor* | | | | | |

| | |
|---|---|
| *I was not pleased by the green spaces close to my house during the COVID-19 crisis* | |
| *Visiting the green spaces close to my house contributed to my well-being during the COVID-19 crisis* | |
| *I was afraid to visit green spaces during the COVID-19 crisis* | |
| *I enjoyed hearing more birds than usual during the COVID-19 crisis* | |
| *I noticed more species around my house during the COVID-19 crisis* | |

For each of the following, please rate the extent to which you agree with each statement, using the scale from 1 (*Strongly disagree*) to 5 (*Strongly agree*). Please respond as you really feel, rather than how you think "most people" feel.

| | 1<br>Strongly disagree | 2 | 3 | 4 | 5<br>Strongly agree |
|---|---|---|---|---|---|
| *My ideal vacation spot would be a remote, wilderness area* | | | | | |
| *I always think about how my actions affect the environment* | | | | | |
| *My connection to nature and the environment is a part of my spirituality* | | | | | |
| *I take notice of wildlife wherever I am* | | | | | |
| *My relationship to nature is an important part of who I am* | | | | | |
| *I feel very connected to all living things and the earth* | | | | | |

**How much time on average it takes you to get from your home to the nearest green space or natural area?**

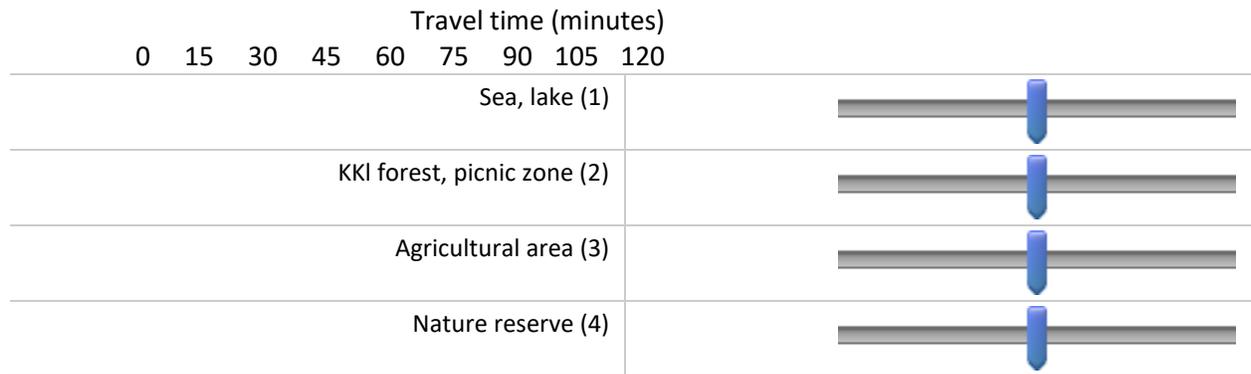

How much time on average it takes you to get from your home to the nearest urban green space?

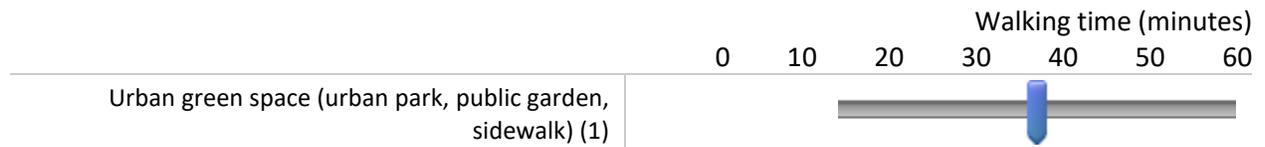

**Try to evaluate how many days on average did you go out to travel in a natural area or green spaces during the past month?**

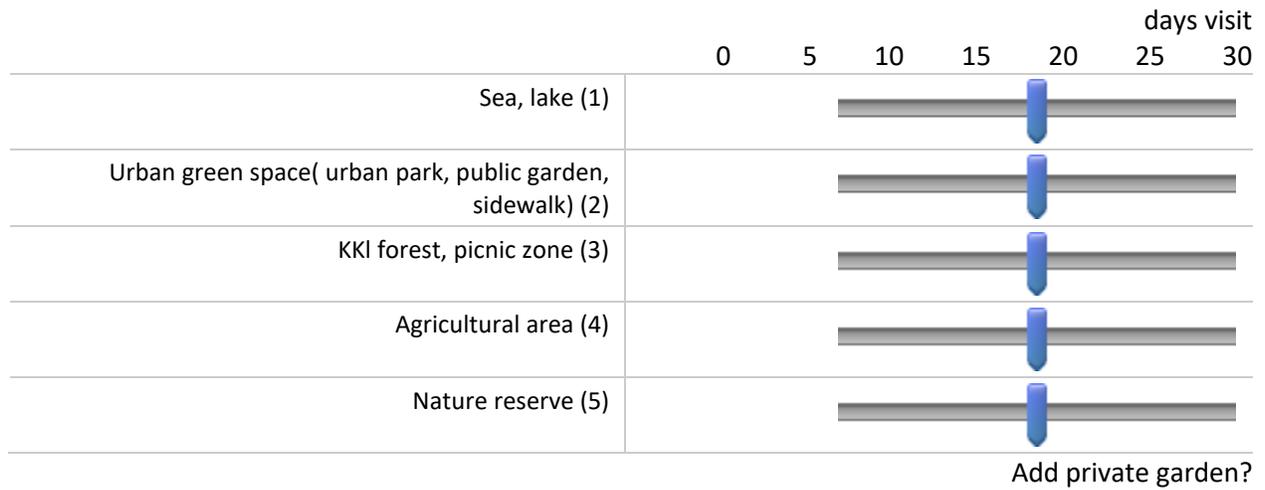

Add private garden?

**Try to evaluate how much time on average did you spend in green space or natural areas (each visit during the past month?)**
**120 minutes= two hours ; 240 minutes= 4 hours ; 300 minutes= 5 hours; 420 minutes= seven hours**

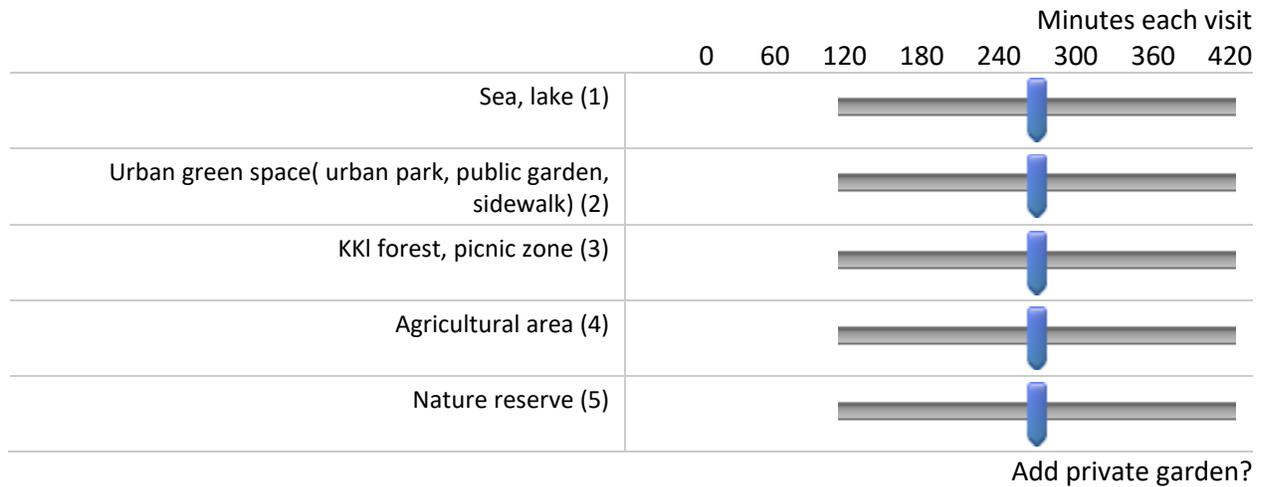

Add private garden?

**Here are several activities, please mark on each activity, to what extant did you preform it when you went out to visit natural area or green space during the past month?**

|  | Never | | | | | | | Each visit |
|---|---|---|---|---|---|---|---|---|
|  | 0 | 1 | 2 | 3 | 4 | 5 | 6 | 7 |
| Watching animals (1) | | | | | | ▮ | | |
| Observing flowers (2) | | | | | | ▮ | | |
| Bathe (3) | | | | | | ▮ | | |
| Picking flowers (4) | | | | | | ▮ | | |
| Go on camping (5) | | | | | | ▮ | | |
| Take pictures of nature (6) | | | | | | ▮ | | |
| Listen to bird chirp (7) | | | | | | ▮ | | |
| Smell flowers (8) | | | | | | ▮ | | |
| Go on a hike (9) | | | | | | ▮ | | |
| Feeding animals (10) | | | | | | ▮ | | |
| Picnic (11) | | | | | | ▮ | | |
| Jeep tour (12) | | | | | | ▮ | | |

How many days did you carry out physical activity for 30 minutes or more during the past week?

………………….

The questions in this scale ask you about your feelings and thoughts during the last month. In each case, you will be asked to indicate how often you felt or thought a certain way (From 0 Never, to 4 very often)

|  | 0 | 1 | 2 | 3 | 4 |
|---|---|---|---|---|---|
|  | Never |  |  |  | Very often |
| *In the last month, how often have you been upset because of something that happened unexpectedly?* |  |  |  |  |  |
| *In the last month, how often have you felt that you were unable to control the important things in your life?* |  |  |  |  |  |
| *In the last month, how often have you felt nervous and "stressed"?* |  |  |  |  |  |
| *In the last month, how often have you felt confident about your ability to handle your personal problems?* |  |  |  |  |  |
| *In the last month, how often have you felt that things were going your way?* |  |  |  |  |  |
| *In the last month, how often have you found that you could not cope with all the things that you had to do?* |  |  |  |  |  |
| *In the last month, how often have you been able to control irritations in your life?* |  |  |  |  |  |
| *In the last month, how often have you felt that you were on top of things?* |  |  |  |  |  |
| *In the last month, how often have you been angered because of things that were outside of your control?* |  |  |  |  |  |
| *In the last month, how often have you felt difficulties were piling up so high that you could not overcome them?* |  |  |  |  |  |

Are you currently receiving any treatment for high blood pressure? ☐ Yes  / ☐ No Please read each statement and indicate the extent to which the statement applied to you over the past week. There is no right or wrong answer.

|  | *Did not apply to me at all* |  |  |  | *Applied to me very much or most of the time* |
|---|---|---|---|---|---|
|  | 0 | 1 | 2 | 3 | 4 |
| *I couldn't seem to experience any positive feeling at all* |  |  |  |  |  |
| *I found it difficult to work up the initiative to do things* |  |  |  |  |  |
| *I felt that I had nothing to look forward to* |  |  |  |  |  |
| *I felt downhearted and blue* |  |  |  |  |  |
| *I was unable to become enthusiastic about anything* |  |  |  |  |  |
| *I felt I wasn't worth much as a person* |  |  |  |  |  |
| *I felt that life was meaningless* |  |  |  |  |  |

The following questions ask how satisfied you feel with your life in general, on a scale from 0 (completely dissatisfied) to 4 (completely satisfied).

Please tick one option in each line to indicate your level of satisfaction with each item.

| *Thinking about your own life and personal circumstances, how satisfied are you with:* | *Completely dissatisfied* |  |  |  | *Completely satisfied* |
|---|---|---|---|---|---|
|  | 0 | 1 | 2 | 3 | 4 |
| *Your life as a whole?* |  |  |  |  |  |
| *Your standard of living?* |  |  |  |  |  |
| *Your health?* |  |  |  |  |  |
| *What you are achieving in life?* |  |  |  |  |  |
| *Your personal relationships?* |  |  |  |  |  |
| *How safe you feel?* |  |  |  |  |  |
| *Feeling part of your community?* |  |  |  |  |  |
| *Your future security?* |  |  |  |  |  |
| *Your spirituality or religion?* |  |  |  |  |  |

Now consider how satisfied you feel about the local neighborhood environment in which you live – i.e. the area within a 1-2-kilometer radius of your house.

Please tick one option in each line to indicate your level of satisfaction with each item.

| Thinking about your neighborhood environment, how satisfied are you with: | Completely dissatisfied | | | | Completely satisfied |
|---|---|---|---|---|---|
| | 0 | 1 | 2 | 3 | 4 |
| Your neighborhood environment as a whole? | | | | | |
| The opportunity for rest and relaxation here? | | | | | |
| Feeling that this environment reflects who you are? | | | | | |
| Memorable experiences here? | | | | | |
| Looking forward to spending time here in the future? | | | | | |
| Being able to think about or reflect on personal matters here? | | | | | |
| The advantages of your neighborhood over others? | | | | | |
| Feeling that you belong in this environment? | | | | | |
| Feeling happy in this neighborhood? | | | | | |

To what extent do you agree with each of the following statements?

| | 1 strongly disagree | 2 | 3 | 4 | 5 strongly agree |
|---|---|---|---|---|---|
| The so-called "ecological crisis" facing humankind has been greatly exaggerated | | | | | |
| The earth is like a spaceship with limited room and resources | | | | | |
| If things continue on their present course, we will soon experience a major ecological catastrophe | | | | | |
| The balance of nature is strong enough to cope with impacts of modern industrial nations | | | | | |
| Humans are severely abusing the environment | | | | | |

From now on, how frequently you are you willing to carry out each of these behaviors?

| | 0 Never | 1 | 2 | 3 | 4 very often |
|---|---|---|---|---|---|
| Recycled paper, plastic, metal | | | | | |
| Conserved water or energy in my home | | | | | |
| Bought environmentally friendly and/or energy efficient products | | | | | |
| Made my yard or my land more desirable to wildlife | | | | | |
| Voted to support a policy or regulation that affects the local environment | | | | | |
| Donated money to support local environmental protection | | | | | |
| Recruited others to participate in wildlife recreation activities | | | | | |
| Volunteered to improve wildlife habitat in my community | | | | | |
| Participated as an active member in an environmental group | | | | | |

What is your year of birth? ………………………….

Are you: □ a man  / □ a woman

What is your highest education level:
□ Below High School, □ High School, □ Bachelor or Professional diploma, □ Above Bachelor

Where do you currently live? Please give us your city and street name, and approximate number in the street: City:……………….. Street:……………………. Approximate number in the street:……

How tall are you? …………………….

How much do you weight? …………………….

The average monthly income per household in Israel is 15,000NIS. How do you judge the monthly income of your household?

| Low | | | | | Average | | | | | High |
|---|---|---|---|---|---|---|---|---|---|---|
| 0 | 1 | 2 | 3 | 4 | 5 | 6 | 7 | 8 | 9 | 10 |

## THANK YOU !